\documentclass[%
 reprint,
 amsmath,amssymb,
 aps,
prb,
]{revtex4-2}

\usepackage{graphicx}
\usepackage{dcolumn}
\usepackage{bm}
\usepackage{hyperref}
\usepackage[mathlines]{lineno}
\usepackage{multirow}
\usepackage{xcolor}

\begin{document}


\title{Large ordered moment with strong easy-plane anisotropy and vortex-domain pattern in the kagome ferromagnet Fe$_3$Sn}

\author{Lilian Prodan$^{1,*}$, Donald M. Evans$^1$, Sin\'ead M. Griffin$^{2,3}$, Andreas~\"Ostlin$^4$, Markus Altthaler$^1$, Erik Lysne$^5$,  Irina G. Filippova$^6$, Serghei Shova$^7$, Liviu Chioncel$^4$, Vladimir Tsurkan$^{1,6}$, and Istv\'an K\'ezsm\'arki$^1$ }

\affiliation{$^1$Experimentalphysik V, Center for Electronic Correlations and Magnetism, Institute for Physics, Augsburg University, D-86135 Augsburg, Germany}
\affiliation{$^2$Materials Sciences Division, Lawrence Berkeley
National Laboratory, Berkeley, CA 94720, USA}
\affiliation{$^3$Molecular Foundry, Lawrence Berkeley National Laboratory, Berkeley, CA 94720}
\affiliation{$^4$Theoretische Physik III, Center for Electronic Correlations and Magnetism, Institute for Physics, Augsburg University, D-86135 Augsburg, Germany \\
Augsburg Center for Innovative Technologies (ACIT),D-86135 Augsburg, Germany}
\affiliation{$^5$Department of Materials Science and Engineering, Norwegian University of Science and Technology (NTNU), 7043 Trondheim, Norway}
\affiliation {$^6$Institute of Applied Physics, MD 2028, Chișinău, R. Moldova}
\affiliation{$^7$Department of Inorganic Polymers, “Petru Poni” Institute of Macromolecular Chemistry, Romanian Academy, 700487 Iasi, Romania}





\date{\today}

\begin{abstract}
We report the structural and magnetic properties of high-quality bulk single crystals of the kagome  ferromagnet Fe$_3$Sn. The dependence of magnetisation on the magnitude and orientation of the external field reveals strong easy-plane type uniaxial magnetic anisotropy, which shows a monotonous increase from $K_1=-0.99\times 10^6 J/m^3$ at 300\,K to $-1.23\times10^6 J/m^3$ at 2\,K. 
Our \textit{ab initio} electronic structure calculations yield the value of total magnetic
moment of about 6.9 $\mu_B$/f.u.
and a magnetocrystalline anisotropy energy density of 0.406\,meV/f.u. ($1.16\times10^6 J/m^3$) both being in good agreement with the experimental values. The self-consistent DFT computations for the components of the spin/orbital moments indicate that the small difference between the saturation magnetisations measured along and perpendicular to the kagome layers results from the subtle balance between the Fe and Sn spin/orbital moments on the different sites. 
In zero field, magnetic force microscopy reveals micrometer-scale magnetic vortices with weakly pinned cores that vanish at $\sim$3\,T applied perpendicular to the kagome plane. Our micromagnetic simulations, using the experimentally determined value of anisotropy, well reproduce the observed vortex-domain structure. The present study, in comparison with the easy-axis ferromagnet Fe$_3$Sn$_2$, shows that varying the stacking of kagome layers provides an efficient control over magnetic anisotropy in this family of Fe-based kagome magnets.

\bigbreak
\textbf{Notes.} \textit{Version 2. This manuscript is under review in Physical Review B.}
\end{abstract}

\maketitle


\section{Introduction}
Magnetic compounds with kagome-lattice arrangement of spins have recently attracted much attention due to their unusual magnetic and electronic properties related to the specific topology of their electronic band structures \cite{bale.10,ma.je.14, bo.na.19,yi.zh.19,ch.to.21}. Recent theoretical and experimental studies have demonstrated that the existence of flat bands, nodal points and nodal lines appearing close to the Fermi energy significantly affect magnetic, magneto-transport and magneto-optical properties of kagome materials~\cite{Zhang2021}.
In particular, fascinating physical phenomena like  giant anomalous and topological Hall effects, giant Nernst effect, topological superconductivity, magnetic spin chirality and skyrmion bubbles have been reported for large group of kagome materials like van der Waals $A$V$_3$Sb$_5$ ($A$ = K, Cs, Rb) \cite{Yang2020, Ortiz2020}, rare earth based $ReT$$_6$Sn$_6$ ($Re$ = Gd, Tm, Tb, Y; $T$ = Mn,V) ~\cite{Ma2021,Peng2021}, magnetic Weyl-semimetal Co$_3$Sn$_2$S$_2$ ~\cite{li.su.18,gu.vi.19}, and binary metals $T$$_x$Sn$_y$ ($T$ = Fe, Mn, Co; $x$:$y$ = 1:1, 3:2, 3:1) ~\cite{ ye.ka.18, na.ki.15, ka.ye.20, ha.li.22, ch.su.22,ho.re.17}. The coexistence of these effects provide an exceptional platform to study electronic band topology and its interplay with magnetic spin and orbital effects.

Here we focus on the magnetic properties of Fe$_3$Sn  quasi-2D kagome magnet, investigated by anisotropy measurements, \textit{ab initio} calculations and imaging of the magnetic domain pattern. Fe$_3$Sn is an itinerant ferromagnet, a member of a family of iron stannides with the general formula Fe$_x$Sn$_y$. Our particular interest is on compounds with the kagome-lattice arrangement, namely compounds with $x:y$ = 1:1; 3:1; 3:2. Depending on the ratio of $x:y$, i.e. the stacking of kagome layers, these compounds realize various hexagonal magnetic space groups and different magnetic ground state. For example, FeSn crystallises in the $P6/mmm$ symmetry and below 365\,K it is an easy-plane antiferromagnet with ferromagnetic arrangement of spins within each kagome layer ~\cite{gi.ni.06,ha.er.75, Sales2019}. Fe$_3$Sn has a structure of $P6_3/mmc$ symmetry and exhibits ferromagnetic order below 743\,K~\cite{gi.ni.06,ca.ma.78}. The crystal structure of Fe$_3$Sn is shown in Figs. 1(a) and (b). The $ab$ layer consists of a breathing kagome lattice of two sizes of equilateral Fe triangles with Sn atoms in the center of the kagome. The unit cell of Fe$_3$Sn contains two adjacent, laterally displaced kagome layers, separated by half of the lattice constant along the $c$ axis. The structure of Fe$_3$Sn is much simpler than of Fe$_3$Sn$_2$, which realizes a third type of stacking and crystallizes in the $R\overline{3}m$ symmetry. Namely, Fe$_3$Sn$_2$ contains two such basic blocks of Fe$_3$Sn bilayers stacked along the $c$ axis and separated by a honeycomb layer of Sn atoms \cite{gi.ni.06,ca.ma.78}. In contrast to Fe$_3$Sn, Fe$_3$Sn$_2$ is an  easy-axis ferromagnet with a Curie temperature of 612\,K~\cite{tr.bo.70}, which shows skyrmion bubbles at room temperature~\cite{ho.re.17,al.ly.21}. However, this compound shows a spin reorientation below room temperature, when the easy axis of magnetisation gets tilted towards the kagome plane~\cite{ca.ma.78,ku.ra.81,Heritage.2020,He.2021}.  

Physical properties of Fe$_3$Sn are investigated predomimantly on polycrystalline samples synthesized by solid state reaction or by arc-melting of metalic Fe and Sn in argon atmosphere. Early M{\"o}ssbauer studies of Trumpy et al.~\cite{tr.bo.70} and Dj\'ega-Mariadassou et al.~\cite{dj.le.66} revealed a ferromagnetic behavior of Fe$_3$Sn and concluded that the spin moment is oriented along the $c$ axis. In contrast, recent analysis of magnetic properties of Fe$_3$Sn by Sales et al.~\cite{sa.sa.14}, performed on field-oriented polycrystalline powder, implied predominantly in-plane orientation of the magnetisation with the magnitude of easy-plane anisotropy $K_1=-1.8 MJ/m^3$ at 300\,K. The easy-plane character of the anisotropy was further supported by magnetisation studies of Fayyazi et al.~\cite{fa.sk.19}, that was performed on an oriented plaque of very small crystals with a mass of 10\,$\mu$g, obtained by reactive flux technique. However, this work reported an unusual temperature dependence of the anisotropy field, namely its decrease from 200\,K to 10\,K.

The precise quantification of magnetic anisotropy is the prerequisite to understand the diversity of complex spin textures, for which kagome magnets offer a fertile ground~\cite{wa.ar.22}.
Moreover, the magnetic anisotropy in interplay with the Dzyaloshinskii-Moriya interaction, competing and frustrated exchanges, or dipole-dipole interactions, can affect the properties of mesoscale spin textures, including conventional domain walls, vortices,  and skyrmions \cite{Bogdanov1980, Yu2012, Okubo2012, Montoya2017, Tang2020, Preissinger2021}. As a prime example, the competition between easy-axis magnetic anisotropy and dipole-dipole interaction can stabilize magnetic bubbles and skyrmions in finite magnetic fields \cite{Yu2012,al.ly.21, ho.re.17, Yu2016}. On the other hand, the in-plane magnetized systems, e.g., magnets with easy-plane anisotropy, can develop magnetic vortices and anti-vortices \cite{Gouva1989, Shigeto2002, Chmiel2018, Gao2019, Zhang2020}. 

Although Fe$_x$Sn$_y$ compounds are known for long, the microscopic origin and the precise value of exchange interactions and magnetic anisotropy constants is still unsettled in these compounds, even the spin arrangement is under debate in some of these materials. From the experimental point of view, the situation is complicated by large variation of sample quality, leading to inconsistencies of the data available in literature. A large piece of the data was obtained on polycrystalline samples, while those reported for single crystals show large scattering evidently related to different quality of samples, e.g., homogeneity, deviation from the stoichiometry and impurity content. Therefore, precisely quatifying the magnetic interactions in a single bilayer Fe$_3$Sn is a prerequisite for understanding magnetism on the microscopic level in these kagome magnets.

Here we report detailed magnetometry studies performed on large stoichiometric single crystals grown by the chemical transport reactions. The appropriate size of the crystals (several mg) allowed us to accurately determine their orientation prior to the magnetisation measurements along the different crystallographic directions. Moreover, our angular-dependent magnetisation measurements, performed in different fields and at various temperatures, allowed a highly reliable quantification of the anisotropy constants of Fe$_3$Sn. The experimental studies were complemented by \textit{ab initio} electronic structure calculations to determine the magnetic properties in the ground state (magnetic moments on Fe and Sn as well as anisotropy energies). 

The spin/orbital moments were restricted to specific orientations and by including the spin-orbit coupling, calculations revealed easy-plane magnetic anisotropy.
The computed uniaxial magnetocrystalline anisotropy energy, which arises from the collective effect of the crystal structure and spin-orbit coupling, is in excellent agreement with the experimentally determined value. Furthermore, a slight departure from the collinear magnetic configuration is predicted due to tilting of the orbital-magnetic moment of Fe with respect to its spin moment. 
Finally, our MFM study reveals the formation of a complex domain pattern built of in-plane flux enclosure domains, i.e. magnetic vortices, on the micrometer scale. This vortex-domain pattern is well reproduced by our micromagnetic simulations using the magnetic interaction parameters determined experimentally and theoretically, further supporting the high reliability of these parameters.

\section{METHODS}

\textbf{Crystal growth.}   Polycrystalline Fe$_3$Sn was prepared by solid state reactions of high purity
elements of Fe (99.99$\%$) and Sn (99.995$\%$). 
A stoichiometric amount of Fe and Sn has been mixed and sealed in quartz ampule evacuated to $10^{-3}$\,mbar. The mixture was annealed for four days at 780\,$^\circ$C followed by quenching in ice water. In order to achieve the full reaction and homogeneity two sintering cycles were performed. The phase purity after each sintering was checked by x-ray powder diffraction. Single crystals of Fe$_3$Sn have been grown by the chemical transport reactions method using the polycrystalline powder as starting material and I$_2$ as the transport agent. The growth was performed in exothermic conditions in a two-zone horizontal furnace with the temperature gradient of 50\,$^\circ$C in a temperature range of 850-800\,$^\circ$C. Crystals up to 1.5\,mm size have been obtained after 6 weeks of transport. The chemical composition of single crystals was analyzed by the ZEISS Crossbeam 550, using energy dispersive x-ray spectroscopy (EDS) technique. 

\textbf{Single-crystal x-ray diffraction.} The single-crystal x-ray diffraction measurement was carried out with a Rigaku Oxford-Diffraction XCALIBUR E CCD diffractometer equipped with graphite-monochromated MoK$\alpha$ radiation. The data were corrected for the Lorentz and polarization effects and for the absorption by multi-scan empirical absorption correction methods. The structure was refined by the full matrix least-squares method based on $F^2$ with anisotropic displacement parameters. The unit cell determination and data integration were carried out using the CrysAlis package of Oxford Diffraction \cite{CrysAlisPro.2015}. All calculations were carried out by the programs SHELXL2014 \cite{Sheldrick.2015} and the WinGX software \cite{Farrugia.1999}. 

\textbf{Magnetic properties.} Magnetic properties were measured by a SQUID magnetometer (MPMS 3, Quantum Design) in the temperature range of 1.8--900\,K and magnetic fields up to 7\,T. The angular-dependent magnetization measurements have been performed both within $ab$ and $ac$ planes on cylindrical sample with the $ac$ plane as the basal plane of the cylinder. The measurements in fields up to 14\,T were performed with a vibrating-sample magnetometer using a physical properties measurement system (PPMS, Quantum Design).

\textbf{Computational Details.}
Density Functional Theory (DFT) calculations were performed with the Vienna \textit{Ab initio} Simulation Package (vasp) \cite{Kresse1993,Kresse1994,Kresse1996} using projector augmented wave (PAW) pseudo-potentials~\cite{Blo,Kresse1999}. We treated the Fe(3d, 4s) and Sn(5s, 5p) electrons as valence. Plane waves were expanded to an energy cutoff of 800\,eV, and a Monkhorst-Pack Gamma-centered k-point grid of $10\times 10\times 10$ for structural optimizations and $20 \times 20 \times 20$ for accurate total energy calculations to obtain the magnetic energetics. A smearing value of 0.02 eV was used for all calculations. The PBEsol functional was used throughout \cite{PBE1}. Spin-orbit coupling was included self-consistently as implemented in VASP. The electronic convergence criteria set to 10$^{-8}$\,eV and the force convergence criteria set to 0.002\,eV/\AA. Our reference structure was taken from the Materials Project (mp-20883)~\cite{MP, Nial:1947, Buschow_et_al:1983, Basile_et_al:1969}.

ELK~\cite{elk} computation has been performed for the a equally large $k$-mesh
($21 \times 21 \times 21 $), and the PBE-GGA functional was used with the experimental lattice parameters given in Table I. For the FPLAPW calculations the plane-wave cutoff $K_{max}$ was set so that $R_{mt}K_{max} = 7.0$, where $R_{mt}$ is the smallest muffin-tin radius.

\textbf{MFM imaging. } The magnetic domain pattern was imaged using a low-temperature attocube attoAFM I, in MFM mode, equipped with a superconducting magnet. The magnetic texture on an as-grown \textit{ab} surface was recorded via a phase-sensitive feedback loop that records changes in the resonance frequency, which are proportional to the gradient of the stray magnetic field of the sample along the magnetisation of the MFM tip, that is normal to the surface.  The magnetic tips used were PPP-MFMR probes from NanoSensor with the magnetic moment of {$5.09\times{10^{-13}\, A/m^2}$}. All data was collected at 100\,K for experimental reasons, such as the thermal stability of the equipment.

\textbf{Micromagnetic simulations. } Simulations were carried out using MuMax3 v3.10 \cite{Vansteenkiste.2014}. Here, the microstructure is found by minimizing a total energy made up of terms representing, Heisenberg exchange, first-order uniaxial anisotropy, Zeeman and demagnetisation energy. Practically this is done by solving the equation
\begin{align}
 \varepsilon=\int_{V_S}[A_{ex}(\Delta m)^2-K_u m_z^2 + M_s B_{ext} m -
\frac{1}{2}M_s B_{dem} m]\textit{d}r 
\end{align}

Where the reduced magnetisation, m(x,y,z), is 
the magnetisation M(x,y,z) divided by the saturation magnetization, $M_s$, within a sample volume $V_S$, where the magnetization makes up a continuous vector field. The exchange stiffness constant, the first order uniaxial anisotropy constant, the external magnetic field and the demagnetisation field are represented by $A_{ex}$, $K_u$, $B_{ext}$, $B_{dem}$, respectively. Periodic boundary conditions were not necessary; the simulation was initiated from a random state, and easy-plane anisotropy (0, 0, 1) was used. The mesh was $4\times4\times4$\,nm cells, and the geometry was $2048\times1024\times256$\,nm. Note, that MuMax3 is able to calculate the MFM response as it knows the orientation of the magnetisation, and therefore the stray fields, and the MFM response comes from the interactions between the MFM tip and the gradient of the stray fields.

\section{RESULTS AND DISCUSSIONS} 
\label{sec:results}
\subsection{Structure, magnetic properties and anisotropy}

Typical as-grown Fe$_3$Sn crystals used in our study are shown in the inset of Fig. \ref{fig:fig2} (a). The energy-dispersive x-ray analysis found an excess of Fe $\lesssim$1\,$\%$, indicating  that the stoichiometry Fe$_3$Sn is close to ideal. Single-crystal x-ray diffraction study  revealed the hexagonal $P6_3/mmc$ ($\#194$) space group symmetry and showed no traces of impurity phases. The calculated cell parameters $a$ and $c$ are close to those reported for polycrystalline Fe$_3$Sn~\cite{gi.ni.06,sa.sa.14}. The structural parameters obtained from the single-crystal refinement are summarised in Table \cite{tab:table1}.

The magnetisation measurements were carried out on a single crystal with a size of $0.7\times0.6\times0.5\,mm^3$. The temperature-dependent magnetisation in 1\,T applied along the $a$ axis exhibits a steep increase below 750 K\,indicating the onset of ferromagnetic order, as shown in Fig. \ref{fig:fig2} (a). The Curie temperature of $T_C =  705$\,K was obtained as the location of the maximum in the temperature derivative of the magnetisation, and is close to previously reported data for polycrystalline samples~\cite{gi.ni.06,ca.ma.78}.

\begin{figure} [t!]
    \includegraphics[scale=0.55]{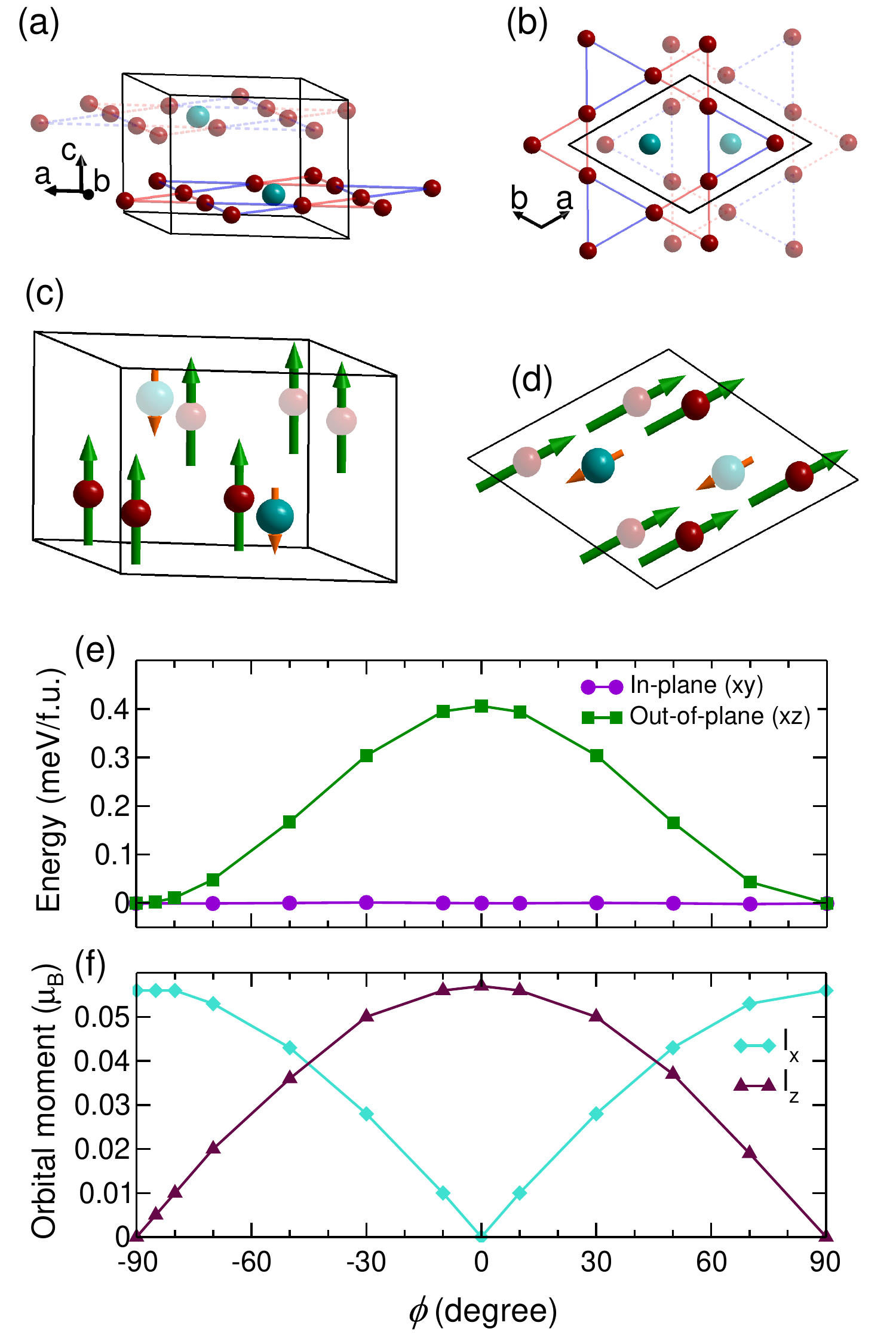}
    \caption{(a) Schematic representation of the crystal structure of layered Fe$_3$Sn. Red and green spheres represent Fe and Sn ions, respectively. Transparent and bold atoms are from different planes. (b) Breathing kagome lattices of equilateral Fe triangles of two distinct sizes (marked by blue and red lines) in the $ab$ plane, as seen when viewed along the $c$ axis. (c)/(d) Magnetic structure with moment parallel/perpendicular to the $c$ axis. Green and orange arrows show Fe and Sn spin moments, respectively. (e) Calculated magnetocrystalline anisotropy energy for rotation of Fe spin within the $ab$ plane (purple) and the $ac$ plane (green), assuming a ferromagnetic ordering. In the former no angular dependence is observed, while in the latter case the energy has a maximum for spins pointing along the $c$ axis ($\phi=0$). (f) Calculated orbital moment magnitude of Fe for in-plane ($l_x$) and out-of-plane ($l_z$) projections as a function of Fe spin tilt angle.}
    \label{fig:fig1}
\end{figure}

Figure \ref{fig:fig2}(b) shows the field-dependent magnetisation curves, $M(H)$, measured at 2\,K for magnetic fields applied along three orthogonal directions. Two directions are in the basal $ab$ plane, $[\overline{1} 2 \overline{1}0]$ and $[10\overline{1}0]$ being parallel and perpendicular to the $a$ axis, respectively, and the hexagonal $c$ axis (equivalently $[0001]$) is chosen as the third direction. The magnetisation within the $ab$ plane reaches the saturation already at $\sim$1\,T. We found that the saturation is reached in a slightly higher field for $H || [10\overline{1}0]$ than for $H || [\overline{1} 2 \overline{1}0]$, indicating that the $a$ axis is the easy axis of the magnetisation. For fields along the $c$ axis, the saturation takes place at higher fields, above 3\,T. 
The uniaxial magnetocrystalline anisotropy within the $ac$ plane was calculated from the hysteresis curves using an “area method” ~\cite{cu.gr.09}, considering the difference of the integrals $\int_0^{M_s}H_i dM$ for in-plane ($a$ axis) and out-of-plane ($c$ axis) field orientations. Here $H_i$ is the internal field determined as the difference between the applied magnetic field $H$ and the demagnetizing field $DM$, where $D$ is the demagnetisation coefficient. We obtained the value of $K_u=-1.27\times 10^6 J/m^3$ at 2\,K for the uniaxial anisotropy constant, which decreases to $K_u$=-0.97$\times 10^6$ $J/m^3$ at room temperature. A similar approach was used to calculate the sixth-order anisotropy in the $ab$ plane, that led to the value of $2.3 \times 10^4 J/m^3$ at 2\,K, which decreases to $1.8 \times 10^4 J/m^3$ at 300\,K. The anisotropy in the basal plane is nearly two orders of magnitude weaker than uniaxial anisotropy.

The inset in Fig.\ref{fig:fig2}(b) shows the temperature dependence of the magnetisation measured in 7\,T. The value of the saturation moment is 2.27\,$\mu_B$/Fe at 2\,K for field in the $ab$ plane, which is $\sim$2\,\% larger than the saturation value along the $c$ axis. Although this difference is rather small it cannot originate from demagnetisation effects, since the sample has a circular shape in the $ac$ plane. Additional measurements, not shown here, show this difference is present up to 14\,T.

 \vspace{3mm}
 
\begin{table}[t]
\caption{\label{tab:table1} Structural refinement details and crystal data for Fe$_3$Sn determined through single-crystal x-ray diffraction.}
    \begin{tabular}{lllll}
\hline
\hline
 &&Refined empirical formula   && Fe$_3$Sn    \\
 \hline
  &&Space group && $P6_3/mmc$ (No. 194)\\
&&$a$ ({\AA})&& 5.4604(3)   \\
&&$c$ ({\AA})&& 4.3458(3)   \\
&&Volume ({\AA$^3$}) && 112.215(15)  \\
&&Z, D$_{calc}$ (g/cm$^3$) && 2, 8.472  \\
&&$\mu$  (mm$^{-1}$) && 29.551    \\
&&$\Theta$   range ($^\circ$)   && 4.310 - 29.014     \\
&&Reflections collected/unique  && 577/68  [R$_{int}$ = 0.0467)]  \\
&&No. variables && 8  \\
&&Goodness-of-fit on $F^2$ &&1.001 \\
&&Extinction coefficient     && 0.174(9)  \\
&&R$_1$, wR$_2$ (all data)     && 0.0143, 0.0352 \\
\hline
    \end{tabular}

\begin{tabular}{cccccccccccccc}

Atom && Position && $x$ &&& $y$ &&& $z$  &&U$_{eq}$\\
\hline
Fe	&&6h &&0.1544(1) &&&0.3088(2)  &&& 0.2500  && 7(1)\\
\hline
Sn	&&2d &&0.6667 &&& 0.3333 &&& 0.2500  && 5(1)\\ 
                  
\hline
\label{tab:I}
\end{tabular}
\end{table}[t]

$^*$U$_{eq}$ is defined as one third of the trace of the orthogonalized $U_{ij}$ tensor.
\vspace{3mm}

\begin{figure}[t]
    \includegraphics[scale=0.48]{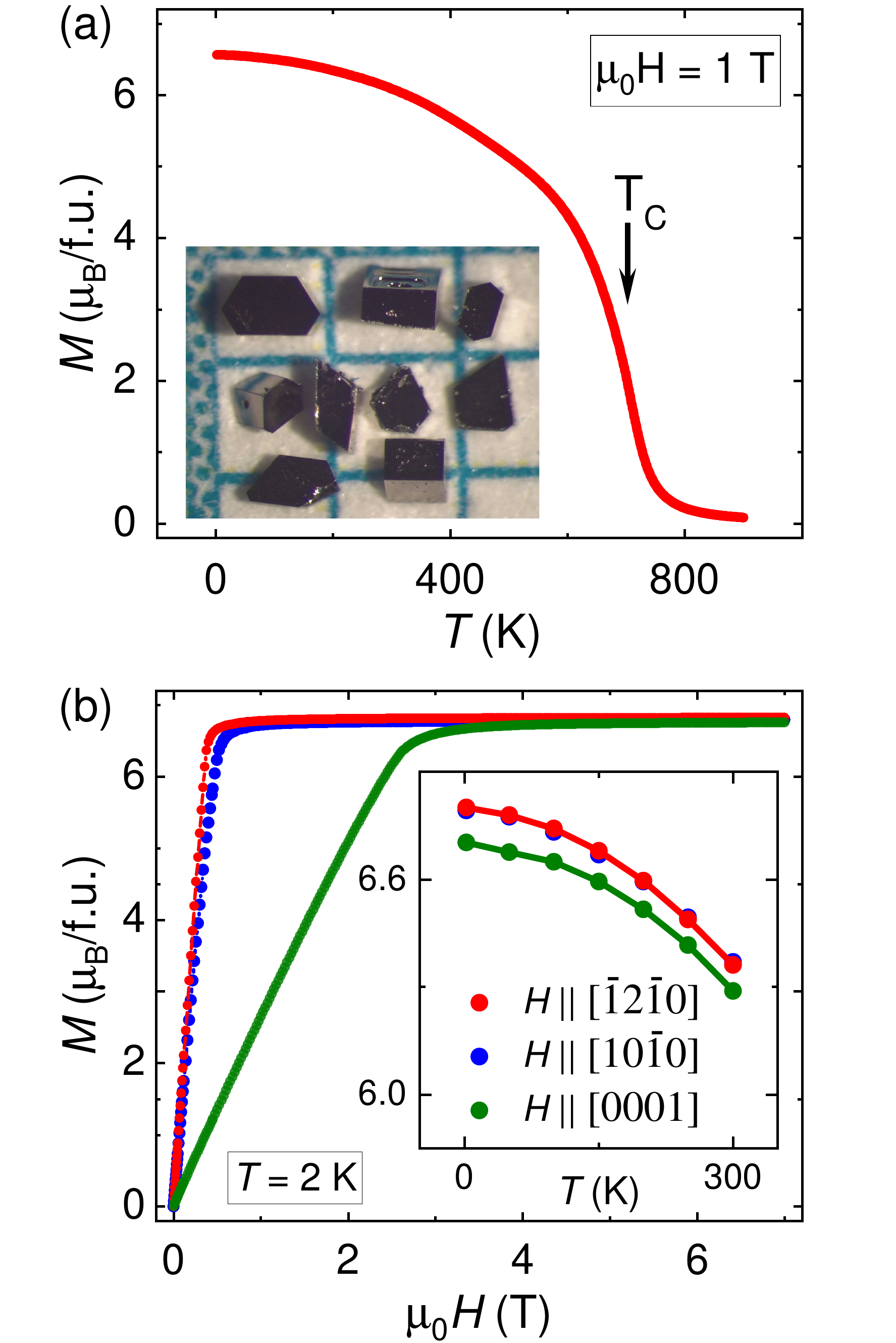}
    \caption{ (a) The temperature-dependent magnetisation of Fe$_3$Sn single crystal measured in 1\,T applied along $a$ axis. The inset shows as-grown Fe$_3$Sn single crystals on a mm-scale paper.(b) Magnetisation curves measured at 2\,K in magnetic fields applied in two directions in the $ab$ plane---parallel to the $a$ axis or $[\overline{1} 2 \overline{1}0]$ and perpendicular to the $a$ axis or $[10\overline{1}0]$, and with fields along the $c$ axis or $[0001]$. The inset shows the temperature dependence of the magnetisation in 7\,T for these three directions. 
     }
    \label{fig:fig2}
\end{figure}

\begin{figure}[t]
  \centering
    \includegraphics[scale=0.63]{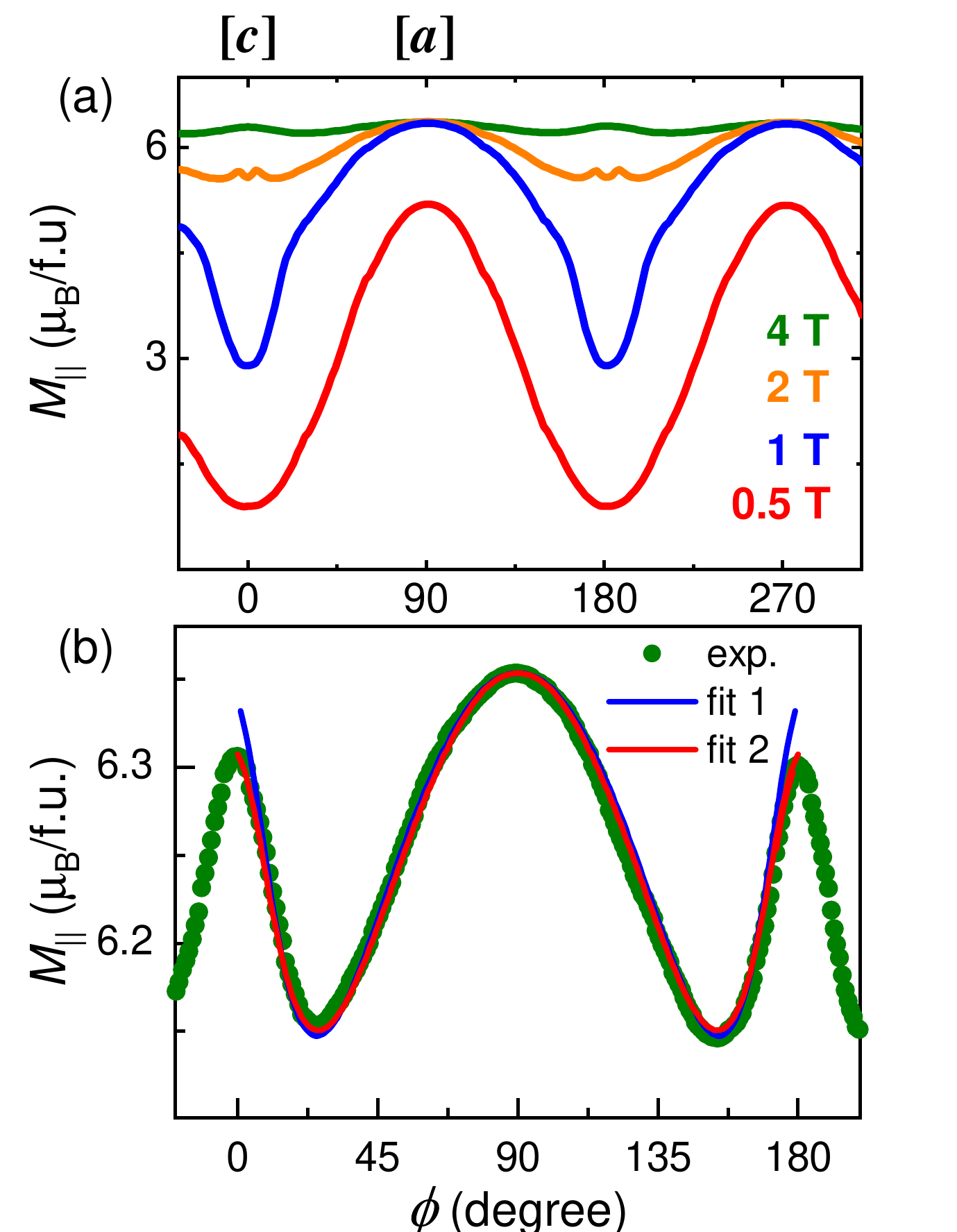}
    \caption{ (a) Angular dependence of the magnetisation measured in different fields at 300\,K. The magnetisation component parallel to the field is detected.   (b) Experimental data (open circles) and fit to the experimental data (solid lines) in 4\,T at 300\,K. }
    \label{fig:fig3}
\end{figure}

The angular dependence of the magnetisation $M(\phi)$ within the $ac$-plane, measured in different fields at 300\,K, is shown in Figure \ref{fig:fig3}(b). Here $\phi$ is the angle spanned by the magnetic field and the $c$ axis (see inset of Fig.~\ref{fig:fig3}(b)). 
As a general trend, the modulation of $M(\phi)$ is suppressed with increasing magnetic field. In addition, there is a change in its functional form. While $M(\phi)$ exhibits a minimum and a maximum in low fields parallel and perpendicular to the $c$ axis, respectively, upon saturation a local maximum develops also for fields applied along the $c$ axis. To avoid complications due to the presence of multiple magnetic domains, we determined the anisotropy based on $M(\phi)$ curves measured in the fully field polarized state, as shown in Fig.~\ref{fig:fig3}(b) for the 4\,T curve. The angular dependence of the magnetisation was fitted based on commonly used phenomenological expression for the magnetic anisotropy energy for a hexagonal ferromagnet  ~\cite{cu.gr.09}
\begin{align}\label{eq:mae}
E_A=K_1 \sin^2 \theta+K_2 \sin^4 \theta+K_3 \sin^6 \theta\cos6 \Psi.    
\end{align}

Here, $K_1$ and $K_2$ are the first- and second-order anisotropy constants, respectively. $\theta$ is the angle spanned by the magnetisation with the $c$ axis, while $\Psi$ denotes the angle between the projection of the magnetisation to the $ab$ plane and the $a$ axis. In our fitting the $K_3$ term is neglected, since our data show that the in-plane anisotropy is less than 2\,\% of the total magnetic anisotropy, as already discussed above.

The fitting formula for calculation of the anisotropy constants $K_1$ and $K_2$ for angular-dependent magnetisation measurements was derived by minimising the total energy per unit volume described as $E=E_A + E_Z + E_D$, where $E_Z=-M_s Hcos(\theta-\phi)$ corresponds to the Zeeman energy, and $E_D=1/2(  M_{||}^2\,D)$ is the magnetostatic energy of a sample in its own field. The formula of fitting of the experimental data is given by 
\begin{align}\label{eq:fit}
\sin2\theta(K_1+2K_2 \sin^2 \theta)=-H_i\sqrt{M_s^2 - M_{||}^2 },
\end{align}
where $M_{||}$ is the projection of $M_s$ on the applied field, which is measured in experiment (see inset of Fig.~\ref{fig:fig3}(b)).

 \begin{figure} [t!]
 \centering
     \includegraphics[scale=0.55]{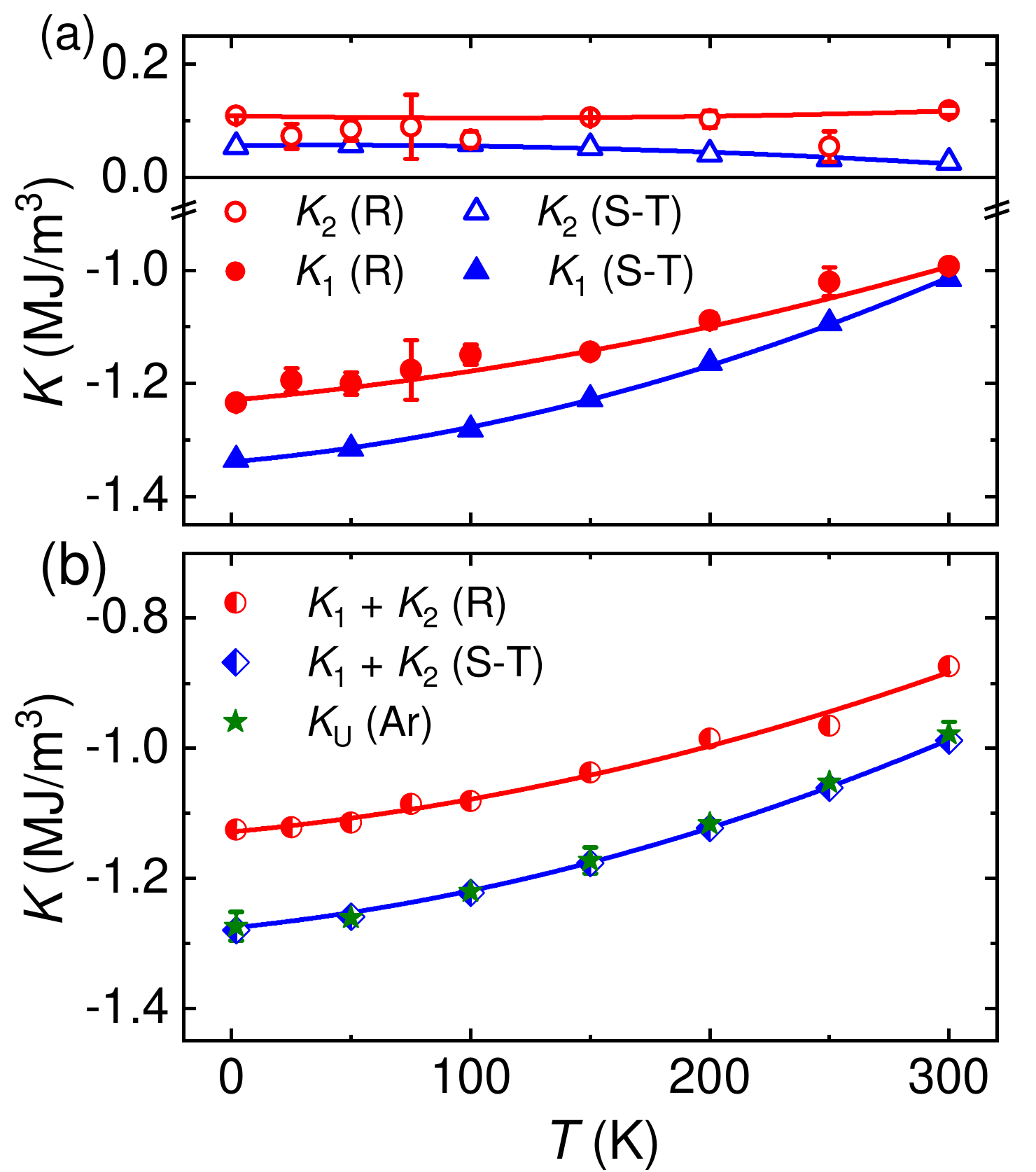}
     \caption{(a)  Temperature-dependent anisotropy constants $K_1$ and $K_2$ as determined from the angular dependence of the magnetisation (circles, marked as R) and by the S-T method (triangles). (b) The uniaxial constant $K_u$ determined by the “area method” (Ar) is plotted together with $K_1$ + $K_2$ taken from panel (a). Lines are guides to the eye.}
     \label{fig:fig4}
 \end{figure}

 The fit to the angular-dependent magnetisation (marked as R method, fit 1 in Fig.~\ref{fig:fig3}(b)) describes well the angular dependence of $M(\phi)$ over the whole angular range except for the immediate vicinity of the $c$ axis, where it overshoots  the experimental data. The experimentally observed difference in $M_s$ for fields parallel and perpendicular to the $c$ axis indicates an anisotropy ($\sim 2\, {\%}$) of the $g$-tensor, i.e. imply peculiar orbital contributions to the magnetisation. (Please notice, if the $g$-tensor anisotropy is completely negligible, the magnetisation should point parallel to the applied field in the high-field limit, hence $\theta$ in Eq.~\ref{eq:mae}
and $\phi$ in Fig.~\ref{fig:fig3}(b) would become identical.)  From the fit to the experimental data we derived the value of the first anisotropy constant $K_1=-1.32\times 10^6 J/m^3$ at 2 K, which is in good agreement with the $K_1=-1.34\times 10^6 J/m^3$ determined by Sucksmith-Thompson (S-T) method (see below) ~\cite{su.th.54}.

To account for the angular dependence of the saturation magnetisation anisotropy we used the approach developed by J. Alameda ${et\, al.}$   and A. S. Bolyachkin ${et\, al.}$ ~\cite{Alameda1980, Bolyachkin2015}. This fit (fit 2 in Fig. \ref{fig:fig3}(b)) improves the description of the experimental data for angles close to the $c$ axis. Figure \ref{fig:fig4} (a) summarizes the temperature dependence of the anisotropy constants $K_1$ and $K_2$, as determined by two independent methods: a) from the fits to the angular dependence of the magnetisation measured in 4\,T and b) calculated from the magnetisation curves along the $c$ axis, following the modified (S-T) method, which considers the anisotropy of the saturation magnetisation~\cite{ Bolyachkin2015}. At 300\,K, both methods yield very close values $K_1=-0.99\times 10^6 J/m^3$ and $-1.01\times 10^6 J/m^3$, respectively. The magnitude of $K_1$ shows a monotonous increase with decreasing temperature in contrast to non-monotonous behavior reported by Fayyazi $et\, al.$~\cite{fa.sk.19}. Moreover, the absolute values of $K_1$ reported in \cite{fa.sk.19} are by about 20\% lower than our experimental values. These discrepancies likely arise due to limitations of proper sample orientation in the former study, which was not an issue in our case due to the larger single crystals. $K_2$ has opposite sign and is lower by approximately one order magnitude than $K_1$. At 300\,K, we obtained $K_2=1.4\times 10^5 J/m^3$ and $2.7\times 10^4 J/m^3$ using the R and S-T methods, respectively.

The reason of the difference in the values of $K_1$ and $K_2$ is unclear and may be related to different data sets for fitting: high-field magnetisation data in the saturated state in  R method and low-field data in S-T method. However, the value of $K_2$ derived from the R method should be taken with precaution because of larger error in determination due to nonlinear extension of the rotator spring and change of its elasticity  with lowering temperature.

To further validate our quantification of the uniaxial anisotropy in Fe$_3$Sn, in Fig. 4(b), we compare the sum of $K_1 + K_2$, as determined using the angular dependent magnetisation and the Sucksmith-Thompson methods, with the overall $K_u$ anisotropy, as obtained using the “area method”. The values obtained by the three methods agree within less than 10\% at all temperatures. 

\subsection{\textit{Ab initio} calculations of magnetic moments and anisotropy}

The origins of magnetocrystalline anisotropy energy (MAE) is well-established for $3d$-transition metals as arising from a combination of the spin-orbit coupling and the crystal-field environment of the magnetic species. Being a ground state property the MAE can in principle be calculated using DFT, however, the small energy scales associated with MAE in $3d$ transition-metal systems make this challenging owing to the tight convergence and sampling needed.  Because of this, we compare two different implementations of DFT - a pseudo-potential code with a plane-wave basis set (VASP), and an all-electron code with LAPW basis (ELK).

DFT computation of Fe$_3$Sn binary system has been recently published~\cite{sa.sa.14, fa.sk.19} and in particular it was found  that the easy magnetic axis lies in the hexagonal plane. Here, we first present our calculations of the magnetic properties using VASP. We find the calculated ground state to be ferromagnetic order with a spin moment of 2.20 $\mu_B$ per Fe, an orbital moment of 0.06 $\mu_B$ per Fe, and a small antiparallel moment of 0.18 $\mu_B$ on the Sn site, giving a total magnetisation of 6.4 $\mu_B$/f.u., consistent with our experimental value of 6.8 $\mu_B$/f.u.. We find no off-plane canting of moments aligned along the easy-plane direction in our VASP calculations, as shown in Fig.\ref{fig:fig1}(f). 
 
To calculate the magnetocrystalline anisotropy, we rotate the spin in the in-plane
and out-of-plane directions, finding the corresponding energy for several angles, as depicted
in Figure \ref{fig:fig1}(e). We find the ground state magnetic order to be ferromagnetic with an easy-plane
anisotropy. In fact, we find the in-plane $({ab})$ spin directionality to be fully degenerate (up to 2 $\mu$eV), in agreement with the negligible sixth-order $ab$-plane anisotropy found experimentally is this study. Our calculated magnetocrystalline anisotropy energy by comparing the energy of the in-plane and out-of-plane spin axes is 0.406 meV/f.u. ($1.16\times10^6 J/m^3$), fully consistent with our experiment. Note that this value is by ca. 20\,\% lower/higher than the theoretical estimate reported by Sales $et\, al.$ ~\cite{sa.sa.14}/Fayyazi $et\, al.$ \cite{fa.sk.19}. In Fig.\ref{fig:fig1}(f) we also plot the variation of the calculated orbital moment on Fe from varying the spin easy-axis in the $({ac})$ direction. For the ground-state spin direction (fully in-plane), we find no out-of-plane contribution to either the spin or orbital moment. However, as the spin axis is rotated in the out-of-plane direction, we find a gradual increase in the out-of-plane orbital moment which is compensated by a decrease in the in-plane orbital moment.

We now discuss our results using ELK. For systems containing $3d$-electrons, the relatively weak SOC can be treated in a Russel-Sounders ($LS$)-coupling scheme,  with well defined spin and orbital quantum numbers. In this case it is possible to define a quantization direction (direction of magnetisation). In practice we have considered the quantization axis (spin moment $\vec{S}$) aligned along the $a$- and $c$- crystallographic directions and performed DFT self-consistent computations for the components of the spin/orbital moments.  
The values of Fe spin/orbital moments are about 2.32/0.07 $\mu_B$ for the considered orientations.
We have noticed, however, that in the self-consistent computation with the moment along the crystallographic $a$-direction the overall (unit cell) moment is slightly increased. The DFT self-consistent configuration for spin moments of Fe and Sn atoms oriented along the $c$/$a$-directions respectively are shown in Figs.~\ref{fig:fig1} (c)/(d).   
In the ferromagnetic ground state the induced spin moment on Sn-sites is of about 0.12 $\mu_B$.
Orbital moments of Sn although existing, have no significant magnitudes ($\le 10^{-3}~\mu_B$). The total magnetic moment per formula unit, as obtained from DFT is 6.9 $\mu_B$/f.u., in good agreement with the experimental value of 6.8 $\mu_B$/f.u.

The DFT moments (on all atoms) computed for the configuration aligned along the $a$-direction are somewhat smaller, 
in particular the decrease in the magnitude of the Sn atoms spin-moments (pointing opposite to the Fe spins) are larger than the corresponding decrease for Fe atoms. 
This is the cause for a slightly larger total (unit cell) moments along the $a$-direction. Thus, the anisotropy of the moments results from the subtle balance between spin and orbital moments on the different sites of the unit cell.

Let us comment further upon the distinction between the anisotropy of magnetic moments and the macroscopic anisotropy energies. 
In systems (containing heavier elements) with lower than cubic symmetry the magnetic moment per atom or the corresponding $g$-tensor is expected to be anisotropic and the spin-orbit interaction might be important also in the absence of a long-range order. Below the ordering temperature, the cooperative alignment of the moments produces a net magnetisation. The energy cost to rotate the magnetisation from the easy axis (lowest energy) into the hard direction is higher at lower temperatures. In the framework of second-order perturbation theory (at zero temperature) the anisotropy energy is proportional to the anisotropy of the orbital magnetic moment.  Note also that for Fe$_3$Sn, for the $c$ direction configuration, the spin and orbital moments align collinearly. At the same time, we found that the orbital moments rotate slightly in the $ab$ plane away from the initial $a$ direction (angular departure of about 7 degrees). Although the bulk measurements presented here are not able to capture the small changes in the orientation and magnitude, recent progress in XMCD allows for a precise determination of spin and orbital moments. Through the XMCD sum rules also the magnitude of the orbital moment and its anisotropy can be measured, note that the relative change in the orbital moment magnitude and direction is important, not the absolute number, which is obviously small.

\begin{figure*} [t!]
 \centering
     \includegraphics[scale=1]{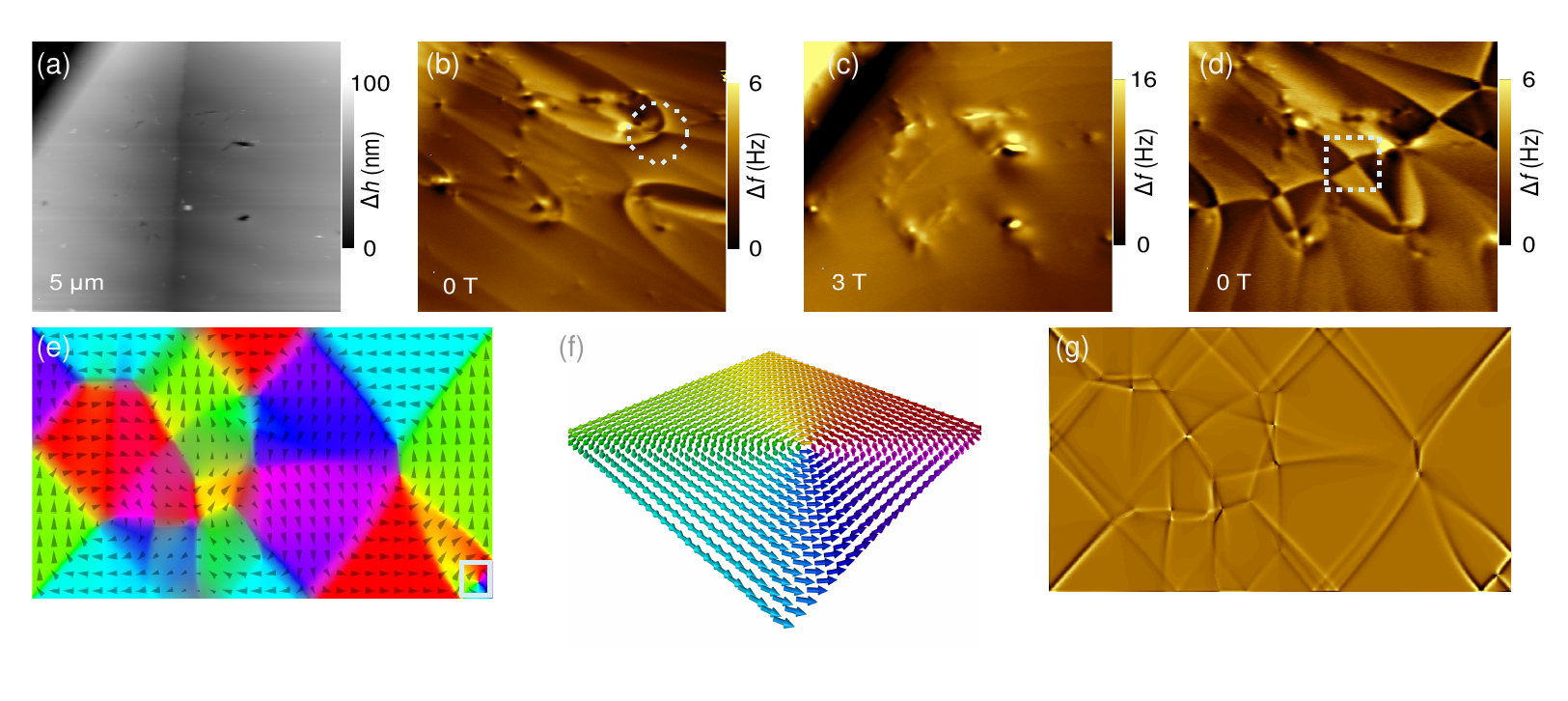}
     \caption{(a) Topographic image collected in contact mode on an as-grown Fe$_3$Sn $ab$ surface, also imaged in the MFM scans in panels (b), (c), and (d). (b)-(d) MFM images, recorded at 100\,K, showing the shift in resonance frequency of the cantilever: Light/dark colors show an increase/decrease in the resonance frequency. (b) Extended domain walls meeting at vortex cores in the zero-field state. (c) Uniform MFM signal, observed except in the vicinity of topographic features, indicate uniform magnetisation in 3\,T. (d) The zero-field state after field treatment hosts a high density of vortices, some of which are much more angular in structure. The difference between the fully relaxed vortex structure (b) and the metastable structure shortly after the removal of a field (d) is highlighted by the dashed circle and square, respectively. (e) Micromagnetic simulations of the local orientation of the magnetisation (gray arrows), viewed down the $c$ axis. The false color code (bottom right) indicates the in-plane orientation of the magnetisation vector. (f) Magnified view of one of the closure structures (vortices) in (e). (g) MFM response calculated for the domain pattern in panel (e).}
     \label{fig:fig5}
 \end{figure*}

\subsection{Imaging and simulations of magnetic vortices}

To move beyond the investigation of bulk properties, we use MFM to image the local magnetic textures on an as-grown $ab$ plane. A contact-mode topography image is given in Fig. \ref{fig:fig5}(a), to show that the surface is smooth enough for the magnetic signal to be collected in a single-pass, constant height mode. While generally flat, there is a change in topography in the top-left of the image and a few point defects which need to be considered when interpreting the MFM data. 
The magnetic microstructure of the sample is represented by the relative shift of the cantilever’s resonance frequency, at a lift height of 150 nm. Figure \ref{fig:fig5}(b) shows a representative MFM image of the virgin zero-field magnetic state. Strikingly this evidences the formation of smooth vortices across the sample. Such vortices are characteristic of systems where the magnetisation is confined to a nearly isotropic plane, either via easy-plane anisotropy or shape anisotropy, the latter being the consequence of dipole-dipole interaction in thin samples.

Following the procedure used for the bulk measurements, we collect MFM images in a 3 T magnetic field, perpendicular to the sample surface. As expected from the bulk measurements, \ref{fig:fig2}(a), Figure \ref{fig:fig5}(c) shows a nearly mono-domain, saturated, state. The remaing contrast predominantly originates at features that correlate with the topography changes seen in Fig. \ref{fig:fig5} (a), suggesting that they are not purely of magnetic origin. The field is removed and a subsequent scan, Fig. \ref{fig:fig5}(d), shows that the vortices have reformed but in different positions and with different geometries. The different positions of the vortices show that the vortex microstructure is not pinned. The more angular geometry of vortex cores, see the high-lighted square in Figure \ref{fig:fig5}(d), are associated with larger stray fields, this likely indicates that the microstructure, at 100 K, has not fully relaxed on the measurement time scale of tens of minutes.

In order to directly connect our images of vortex-antivortex microstructure to our bulk measurements, we use micromagnetic simulation. The bulk data are used as input parameters to compute the expected orientation of the local magnetic moments, as shown in Fig. \ref{fig:fig5}(e). Here, the small arrows represent the nanoscale orientation of the magnetisation vector and the rainbow false colors showing areas of the same orientation. The pattern is typical for an easy-plane magnet, with sets of vortex-antivortex pairs. Figure \ref{fig:fig5}(f) shows one such example of this, with a continuous in-plane rotation of magnetisation.

The connection between the local moment and the stray fields, that our MFM senses, can be counterintuitive, and so we use MuMax3 to calculate the expected MFM response of the microstructure (shown in Fig. \ref{fig:fig5}(e)) which is presented in Fig. \ref{fig:fig5}(g). This is possible once the tip material and lift height are given, as the force on the MFM tip goes as the gradient of the stray field – which is known via the preceding magnetisation calculations. The calculated MFM signal is in excellent qualitative agreement with the observed signal of the zero-field state, e.g. Fig. \ref{fig:fig5}(b) and (d), showing a series of vortex points connected by domain walls. The simulations also show that the contrast at the walls depends on the exact nature of the magnetisation reorientation: some walls have very sharp bright-dark features, while others have smoother transitions; again, as observed in the MFM data. Note, observing in-plane closure domains with an MFM is long established for thin-films, for instance, Ref. \cite{Mamin1989}.

\section{Conclusions}

In conclusion, we have grown high-quality bulk single crystals of  single kagome bilayer ferromagnet Fe$_3$Sn and performed magnetisation and magnetic-force-microscopy studies. The studies revealed a strong uniaxial easy-plane anisotropy characterized by the first-order anisotropy constant $K_1=-0.99\times10^6 J/m^3$ at 300\,K and $-1.23\times10^6 J/m^3$ at 2\,K.  The three independent methods, applied for the calculation of anisotropy constants in our work, provide values that are in good agreement with each other as well as the anisotropy energy obtained from our \textit{ab initio} study.  Moreover, the used angular dependent measurements allowed to evidence the anisotropy in the saturation magnetisation along the  $a$ and the $c$ axes, which we ascribe to orbital contributions. Our DFT calculations predict an induced spin moment on the Sn sites of a similar magnitude to the Fe's orbital moment, the two pointing opposite to each other. All-electron DFT calculations suggest that the orbital-magnetic moments of Fe tilt within the easy plane with respect to the main crystallographic directions, thus a slight departure from the collinear magnetic configuration is obtained. Although the current experimental analysis of the bulk magnetisation data can not distinguish between the orbital collinear and non-collinear configuration within the $ab$-plane, the neutron diffraction measurements might confirm the existence and the magnitude of the induce spin moment on Sn sites. Moreover, XMCD studies have the potential to reveal the predicted non-collinearity of the orbital moments. 

Our MFM study further confirms the system is an easy-plane ferromagnet that contains a rich microstructure of the magnetisation pattern, dominated by (anti)vortices. The experimentally observed MFM contrast is well reproduced by the micromagnetic simulations, using experimentally determined values of the magnetic interaction parameters. 

Overall, our bulk and microscopic experimental studies in combination with \textit{ab initio} and micromagnetic calculations provide a multi-scale, highly reliable approach to quantify magnetic interactions, e.g., magnetic anisotropy and spin arrangement, and to reveal their origins in the kagome bilayer model system Fe$_3$Sn. This is an important step towards understanding complex magnetism emerging in kagome magnets in presence of remarkable spin-orbit effects.

\section*{Acknowledgments}
This work was supported by the Deutsche Forschungsgemeinschaft (DFG) through Transregional Research Collaboration TRR 80 (Augsburg, Munich, and Stuttgart), and by the project ANCD 20.80009.5007.19 (Moldova). 
DME wishes to thank and acknowledge funding by the DFG individual fellowship, number (EV 305/1-1). SMG acknowledge funding by the U.S. Department of Energy, Office of Science, Office of Basic
Energy Sciences, Materials Sciences and Engineering Division under Contract No. DEAC02-05-CH11231 (Nonequilibrium magnetic materials program MSMAG). Computational
resources were provided by the National Energy Research Scientific Computing Center and
the Molecular Foundry, DOE Office of Science User Facilities supported under same contract.



\bibliography{ref}
\end{document}